\documentclass[prb,aps,twocolumn,showpacs]{revtex4-1}
\usepackage{graphicx}
\usepackage{epsfig}
\usepackage{latexsym}
\usepackage{amsmath,amsfonts,amssymb}
\usepackage{color}
\usepackage{array}
\newcommand{\e}{\varepsilon}
\newcommand{\up}{\uparrow}
\newcommand{\down}{\downarrow}
\renewcommand{\>}{\rangle}
\renewcommand{\(}{\left(}
\renewcommand{\)}{\right)}
\renewcommand{\[}{\left[}
\renewcommand{\]}{\right]}
\renewcommand{\v}[1]{\mathbf{#1}} % \v -> vector (bf)
\begin{document}
\title{Majorana End-States in Multi-band Microstructures with Rashba Spin-Orbit Coupling}
\author{Andrew C. Potter and Patrick A. Lee}
\affiliation{ Department of Physics, Massachusetts Institute of
Technology, Cambridge, Massachusetts 02139}

\begin{abstract} A recent work\cite{Potter} demonstrated, for an ideal spinless p+ip superconductor, that Majorana end-states can be realized outside the strict one-dimensional limit, so long as: 1) the sample width does not greatly exceed the superconducting coherence length and 2) an odd number of transverse sub-bands are occupied.  Here we extend this analysis to the case of an effective p+ip superconductor engineered from Rashba spin-orbit coupled surface with induced magnetization and superconductivity, and find a number of new features.  Specifically, we find that finite size quantization allows Majorana end-states even when the chemical potential is outside of the induced Zeeman gap where the bulk material would not be topological.  This is relevant to proposals utilizing semiconducting quantum wires, however, we also find that the bulk energy gap is substantially reduced if the induced magnetization is too large.  We next consider a slightly different geometry, and show that Majorana end-states can be created at the ends of ferromagnetic domains.  Finally, we consider the case of meandering edges and find, surprisingly, that the existence of well-defined transverse sub-bands is not necessary for the formation of robust Majorana end-states.
\end{abstract}
\maketitle

\section{Introduction}
Majorana fermion bound states are expected to exhibit non-Abelian exchange statistics\cite{Ivanov,Read/Green}, and have been proposed as a basis for topological quantum computers which would be protected from decoherence\cite{Kitaev,Nayak}.  Consequently, there is a growing interest in producing Majorana fermions in the laboratory.  Superconductors with $p+ip$ pairing symmetry have long been expected to posses zero-energy Majorana bound states in vortex cores\cite{Ivanov}.  Such $p+ip$ superconductors are thought to naturally occur in triplet paired fermionic superfluids (such as $^3\text{He}$ A  or $\text{Sr}_2\text{RuO}_4$)\cite{He3Ref,Sr2RuO4Ref}, and in the Pfaffian quantum Hall state at $\nu=5/2$\cite{Moore/Read}.  However, despite extensive experimental work on such systems, direct evidence of Majorana fermions remains elusive. 

Recently, the possibility of engineering effective $p+ip$ superconductors in more conventional materials has arisen\cite{Fujimoto,Sau,AliceaSingle,PALee}. A particularly promising class of such proposals involves using Rashba-type spin-orbit coupling in combination with conventional $s$-wave superconductivity to produce an effective $p_x \pm ip_y$ 2D superconductor\cite{Fujimoto,Sau,AliceaSingle}.  Magnetization would then be introduced to remove one of the two components, leaving an effective $p_x + ip_y$ superconductor.    

The practical difficulties of creating and manipulating vortices has led to a renewed interest in the original Kitaev idea\cite{Kitaev}, where Majorana particles are realized as localized states at the ends of a one-dimensional $p_x+ip_y$ superconducting wire.  In addition to being potentially simpler to implement than vortex based proposals, creating Majorana fermions as end-states in quantum wires would allow one to build scalable networks of gates to braid, fuse, and measure many Majorana fermions\cite{AliceaSingle}. A recent work\cite{Potter} demonstrated, for an ideal (spinless) $p+ip$ superconductor, that Majorana end-states can be realized outside the strict one-dimensional limit, so long as: 1) the sample width does not greatly exceed the superconducting coherence length $\xi_0=\pi v_F/\Delta$, and 2) an odd number of transverse sub-bands are occupied.  Furthermore, since Majorana end-states emerge only for an odd number of occupied sub-bands, as chemical potential is swept the system undergoes a sequence of alternating topological phase transitions between phases with and without Majorana end-states. 

In this paper we extend this analysis to the case of spinfull fermions with Rashba spin-orbit coupling and with s-wave pairing and ferromagnetic splitting induced by proximity effect. Because this system is effectively a single species $p+ip$ superconductor for a certain range of parameters, we expect, and indeed find, that the results of Ref. \onlinecite{Potter} still apply.  However, because of the presence of multiple energy scales (Fermi-energy $\e_F$, spin-orbit coupling $\Delta_R$, magnetization $V_z$, and superconductivity $\Delta$), several new features and possibilities emerge.

We begin by analyzing a long narrow strip geometry with hard-wall boundary conditions.  This geometry was previously considered in Ref.  \onlinecite{DasSarmaMultichannel}, which identifies Majorana end-states by computing a topological invariant for a small number of occupied sub-bands.  Here, we extend these results to an arbitrary number of subbands.  We find that, in the parameter regime where the system is effectively a single component $p+ip$ superconductor, Majorana end-states exist when an odd number of transverse sub-bands are occupied. Interestingly for $V_z>\Delta>\Delta_R$, a regime that is relevant to semiconductor materials, finite width quantization allows the Majorana end-states to persist even for $\mu>V_z$ where a two-dimensional sample would be topologically trivial.  This observation allows one to operate at substantially larger carrier density than previously expected\cite{AliceaSingle}, placing less stringent requirements on sample purity.  However, our results also show that the energy gap is substantially reduced if the Zeeman energy gets too large.

We also show that spatially non-homogenous magnetization profiles can be used to produce Majorana end-states at the ends of long rectangular Ferromagnetic domains.  We further demonstrate that structure of Majorana end-states remains largely unchanged for smoothly varying magnetization profiles.  Such geometries are advantageous for microfabricated structures, because the induced magnetization profile will be smooth even if the etched edges of the ferromagnetic material are rough, thus diminishing the impact of edge disorder.  

\begin{figure}[ttt]
\begin{center}
\hspace{-.2in}
\includegraphics[width=3in]{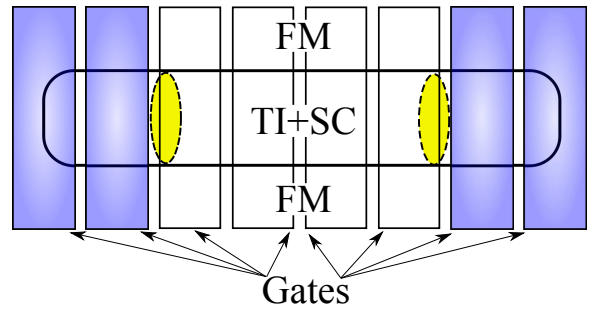}
\end{center}
\vspace{-.2in}
\caption{(Color online) Proposed setup for electrically manipulating Majorana end-states in topological insulator (TI) materials.  A strip of TI with induced superconductivity (labeled TI+SC) is embedded in a ferromagnetic insulator (labeled FM).  Top gates (shown as overlayed rectangles) are used to locally control the number of occupied sub-bands.  Blue shaded gates indicate an even number of sub-bands, demarking a non-topological region, whereas un-shaded gates indicate an odd number of sub-bands, demarking a topological region.  Majorana bound states (shown as yellow blobs) emerge at the boundary between topological and non-topological regions.}
\label{fig:GatedTI}
\end{figure}

In addition, a similar setup could be used to realize end-states in an island of topological insulator (TI) with induced superconductivity (SC), surrounded by ferromagnetic (FM) insulator (see Fig. \ref{fig:GatedTI}).  One can take advantage of the sensitivity of the existence of Majorana end states to the chemical potential by adopting the geometry shown in Figure \ref{fig:GatedTI}.  In this geometry, the Majorana modes could be moved around by selectively applying gate voltages to locally tune the number of occupied sub-bands, thus obviating the need to create and manipulate vortices.

Finally, we consider random, meandering edge geometries in order to address the question of whether or not it is necessary to have well-defined transverse sub-bands in order to produce Majorana end-states.  Suprisingly, we find that the existence of Majorana fermions and the alternating structure of topological phase transitions with chemical potential persists, even when edge variations are large enough that there is no well-defined concept of transverse sub-bands.

The paper is organized as follows: we begin with a short review of the proposed route to engineering an effective single species $p+ip$ superconductor from materials with Rashba spin-orbit coupling, and of the results of Ref. \onlinecite{Potter} for quasi-one-dimensional spinless $p+ip$ superconductors.  We then introduce the tight-binding model which forms the basis of our analysis, and describe a Green's function based method used to treat large system-sizes.  We then describe the resulting analysis of this tight-binding model for hard-wall boundary conditions, ferromagnetic domains, and random non-rectangular samples.

\section{Overview: Topological Superconductivity from Rashba Coupled Surfaces}
In this section we briefly review the proposed route to engineering an effective $p+ip$ superconductor in 2D surfaces with Rashba spin-orbit coupling\cite{Fujimoto,Sau}.  The Hamiltonian for a 2D surface with Rashba coupling is: 
\begin{equation} H_{\text{Rashba}} = \sum_{\mathbf{k}} \[\xi_{k}+\alpha_R\hat{\v{z}}\cdot(\sigma\times\v{k})\]_{\alpha\beta}c_{\v{k}\alpha}^\dagger c_{\v{k}\beta} \end{equation}
where $\xi_k = \frac{k^2}{2m}-\mu$ is the spin-independent band structure, $\mu$ is the chemical potential, $\alpha_R$ is the Rashba coupling strength, $\sigma$ are the spin-$1/2$ Pauli matrices, and $\alpha,\beta = \up\down$ are spin indices.  The Rashba coupling $\alpha_R$ creates two helical bands with energies $\e_{\pm}^{(R)} = \xi_k\pm \alpha_R|k|$ and spin-wavefunctions $\Psi_R^{\pm}=\frac{1}{\sqrt{2}}\begin{pmatrix}\pm e^{i\phi_\v{k}}\\ 1\end{pmatrix}$, where $\phi_\v{k} = \tan^{-1}\(k_x/k_y\)$, which wind counter-clockwise and clockwise respectively.  One can introduce s-wave (spin-singlet) superconductivity by proximity effect:
\begin{equation} H_\Delta = \sum_{\v{k}}\Delta c_{\v{k}\up}^\dagger c_{-\v{k}\down}^\dagger +h.c. \end{equation}
Re--expressing $H_\Delta$ in terms of the helical Rashba surface bands, one finds that that the induced pairing has $p\pm ip$ for $\Psi_R^\pm$ respectively\cite{Fujimoto,Sau}.  

While the Rashba splitting has effectively generated p-wave superconductivity, fermion states on the spin-orbit coupled surface still occur in degenerate time-reversed pairs. Consequently, the system is still topologically trivial and will not exhibit Majorana states.  To break this time-reversal doubling and obtain a topologically non-trivial single-species $p+ip$ superconductor, one can introduce Zeeman splitting term 
\begin{equation} H_{V_z} = \sum_k V_z \(c_{\v{k}\up}^\dagger c_{\v{k}\up}-c_{\v{k}\down}^\dagger c_{\v{k}\down}\)\end{equation} 
through, for example, proximity to a Ferromagnetic insulator.  Such a term modifies the bare ($\Delta=0$) energy bands to 
\begin{equation} \e_{\pm} = \xi_k\pm\sqrt{V_z^2+\alpha_R^2k^2}\end{equation}
$V_z$ also tends to cant the helical bands out of the xy-plane, giving them some component along the z-axis.  Re-expressing $H_\Delta$ in the eigenbasis of both Rashba and Zeeman couplings, one finds that, in addition to $p\pm ip$ pairing $\Delta_{\text{p}}(k) \hat{\v{k}}^\pm\sim  \<c_{k,\pm}c_{-k,\pm}\>$ between fermions both in band $\e_{\pm}$, this canting introduces an s-wave pairing component $\Delta_{\text{s}}(k) \sim \<c_{k,+}c_{-k,-}\>$ between fermions $c_+$ and $c_-$ in bands $\e_{+}$ and $\e_{-}$ respectively where:
\begin{equation} \begin{pmatrix} \Delta_{\text{s}}(\v{k}) \\ \Delta_{\text{p}}(k) \end{pmatrix} = \frac{1}{2\sqrt{V_z^2+\alpha_R^2k^2}}\begin{pmatrix} V_z \\ -\alpha_R k \end{pmatrix} \Delta \label{eq:DeltaSP}\end{equation}
and $\hat{\v{k}}^\pm = \(k_y\pm ik_x\)/k$.  

As discussed in \cite{AliceaSingle}, one has a topological superconductor with potential Majorana bound states so long as $V_z>\Delta$, and so long as $\mu$ lies within the Zeeman gap ($|\mu|<V_z$).  The latter restriction is potentially problematic for realizing the above outlined scheme in semiconductor heterostructures. These structures exhibit small Rashba splittings on the order of $\alpha_R k_F = 2m\alpha_R^2 \sim 10^{-4}eV$ (where $k_F$ is the Fermi momentum for $\mu=0$).  Furthermore, it is also desireable to have Rasbha splitting comparable or larger than Zeeman splitting such that the induced superconductivity has a substantial p-wave component (see Eq.\ref{eq:DeltaSP}).  For very small Rasbha splitting, the conditions $|\mu|<V_z$ and $\alpha_R k_F\gtrsim V_z$ together require low carrier density, making such structures susceptible to disorder.

\section{Majorana End-Modes in Rashba Coupled Structures}
\subsection{Spinless $p+ip$ Case}
Before treating the case of spinfull fermions with Rashba spin-orbit coupling, which will be the focus of this paper, we briefly review some pertinant results from Ref. \onlinecite{Potter} for Majorana end-states in spinless p+ip superconductors.  Ref. \onlinecite{Potter} considers a spinless p+ip superconducting sample of length $L_x$ and width $L_y$.  In the 2D limit where $L_x,L_y \gg \xi_0$ (where $\xi_0 = \pi v_F/\Delta$ is the superconducting coherence length), the sample has a bulk superconducting gap but exhibits a gapless chiral edge mode whose energy is quantized by the finite sample perimeter.  The edge mode wave-function is localized on the system boundary with characteristic length-scale $\xi_0$.  As $L_y$ is decreased below $\xi_0$, the tails of the edge mode wave-functions begin to strongly overlap generating a gap along the length of the sample that scales as $\sim e^{-L_y/\xi_0}$.  

In Ref. \onlinecite{Potter}, it was shown that for an odd number of occupied transverse sub-bands, this $\Delta_{\text{bulk}} \sim e^{-L_y/\xi_0}$ bulk gap stabilizes zero-energy Majorana states isolated on opposite ends of the sample.  Because these end-states occur when an odd number of transverse sub-bands are occupied, as chemical potential $\mu$ is changed, the system undergoes alternating sequence of topological phase transitions between phases with and without Majorana end-states.  

When Majorana end states are spatially well separated, the bulk gap $\Delta_{\text{bulk}}$ exponentially suppresses the probability of an unpaired electron tunneling between any two Majorana states.  This exponential suppression protects the Fermion parity information that is stored non-locally between any two Majoranas throughout any braiding process in which the particles remain well separated.  However, in order to measure the mutual occupation of two Majorana end-states, it is usefull to bring the Majoranas close together, fusing them into a single fermion state which is either occupied or unoccupied.  For example, one can measure the occupation of two Majoranas by coupling them accross a Josephson junction and measuring the sign of the resulting Josephson current\cite{Kitaev}.  When fusing Majoranas for measurement, the important energy scale is not the gap to extended bulk-excitations but rather the so-called mini-gap to localized Fermion end-states.  

\begin{figure}[tth]
\begin{center}
\includegraphics[width=3.5in]{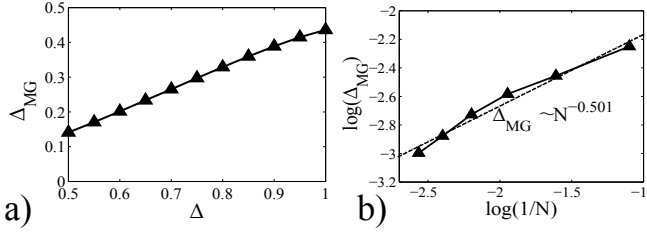}
\end{center}
\vspace{-.1in}
\caption{ Demonstration that the mini-gap $\Delta_{MG}$ protecting the Majorana end-statesof a spinless $p+ip$ superconductor scales as $\Delta/\sqrt{\mathcal{N}}$ rather than $\Delta^2/\e_F$ as is the case for vortex core states.  The simulations were perfomed for a $L_x\times L_y$ strip of the tight-binding model considered in Ref. \onlinecite{Potter} with hopping $t$ and p-wave BCS pairing $\Delta$.  Panel A shows the linear scaling of $\Delta_{MG}$ with $\Delta$ for $L_x = 150$, $L_y=10$, $t=10$ and chemical potential $\mu=-2t$ (corresponding to $\mathcal{N}=5$ filled sub-bands).  Panel B shows the $1/\sqrt{\mathcal{N}}$ scaling of the mini-gap (where $\mathcal{N}$ is the number of occupied subbands), for  $L_x = 250$, $L_y=20$, $t=40$, and $\Delta = 1$.  In Panel B, the dashed line shows the best power-law fit to the simulation data scales as $N^{-0.501}$.}
\label{fig:GapScaling}
\end{figure}

When Majoranas are realized as bound states in Abrikosov vortices, this mini-gap scales as $\Delta_{\text{MG-Vortex}}\sim\Delta^2/\e_F\ll \Delta$.  For simple measurement schemes, this small mini-gap requires working at very low temperature.  More sophisticated measurement schemes that do not require bringing Majoranas close to each other are possible\cite{VortexInterferometry}, however these schemes are comparatively more complex, presenting additional challenges for experiment. In comparison, the mini-gap for Majorana end-states in quasi-one-dimensional wires scales as 
\begin{equation} \Delta_{\text{MG}}\sim \frac{\Delta}{\sqrt{\mathcal{N}}}\end{equation} 
 where $\mathcal{N}$ is the number of occupied sub-bands.  The slow square-root dependence on the number of occupied sub-bands indicates that the mini-gap can be a sizeable fraction of the bulk gap even for many occupied sub-bands.  Figure \ref{fig:GapScaling} demonstrates this $\frac{\Delta}{\sqrt{\mathcal{N}}}$ for exact diagonalization of a quasi-one-dimensional spinless $p+ip$ superconductor.  Panel A and B of Figure \ref{fig:GapScaling}  show the scaling of the mini-gap with $\Delta$ and  $\mathcal{N}$ respectively.  

The parametric scaling of the mini-gap can be understood simply, by considering each sub-band as contributing a Majorana end-state, which overlap spatially, and are coupled to one another by the p-wave pairing gap $\sim \Delta$.  Specifically, the  $\Delta/\sqrt{\mathcal{N}}$ excitation scaling is consistent with that of a random $\mathcal{N}\times\mathcal{N}$ antisymmetric matrix, whose entries are normally distributed with width $\Delta$.

In what follows, we demonstrate that a similar picture holds for effective $p+ip$ superconductors generated from spinfull fermions with Rashba spin-orbit coupling, Zeeman splitting, and induced s-wave pairing.  However, the story is complicated because of the presence of multiple energy scales (Fermi energy, Rashba and Zeeman splittings, and superconducting gap) which gives rise to new features, and allows for a number of generalizations.

\subsection{Tight-Binding Model and Green's Function Method} 
To analyze the structure of Majorana zero-modes in p+ip superconductors engineered from Rasbha coupled structures, we study numerically a discrete square-lattice tight-binding model version of the continuum Hamiltonian:
\begin{eqnarray} && H_{\text{TB}} = H_t+H_{\text{SO}}+H_{\text{FM}}+H_\Delta \nonumber \\
&& H_t =  \sum_{\mathbf{R},\mathbf{d},\alpha} -t \(c_{\mathbf{R}+\mathbf{d},\alpha}^\dagger c_{\mathbf{R},\alpha} + h.c.\)-\mu c_{\mathbf{R}\alpha}^\dagger c_{\mathbf{R}\alpha} \nonumber \\
&& H_{\text{FM}} = \sum_\mathbf{R,\alpha,\beta}  V_z c_{\mathbf{R},\alpha}^\dagger \(\sigma_z\)_{\alpha\beta}c_{\mathbf{R}\beta}+h.c.
\nonumber\\
&& H_{\text{SO}} = \sum_{\mathbf{R},\mathbf{d},\alpha,\beta} -i\alpha_R c_{\mathbf{R}+\mathbf{d},\alpha}^\dagger  \hat{\mathbf{z}}\cdot\(\vec\sigma_{\alpha\beta}\times\mathbf{d}\) c_{\mathbf{R},\beta} + h.c. \nonumber \\
&& H_{\text{SC}} = \sum_\mathbf{R} \Delta c_{\mathbf{R}\up}^\dagger c_{\mathbf{R}\down}^\dagger +h.c. \label{eq:H_TB}\end{eqnarray}
where $\mathbf{R}$ labels lattice sites, $\mathbf{d}\in \{\hat e_x,\hat e_y\}$ is a unit vector connecting nearest neighboring sites, $(\alpha,\beta) \in\{\up\down\}$  are spin-indices, and $\{\sigma_j\}_{j=x,y,z}$ are spin-$1/2$ Pauli matrices.  These terms represent the kinetic hopping energy ($H_t$), induced ferromagnetic Zeeman splitting ($ H_{\text{FM}}$), Rashba spin-orbit coupling ($H_{\text{SO}}$),  and induced spin-singlet pairing ($H_\Delta$) respectively. 

\begin{figure}[tbh]
\begin{center}
\hspace{-.2in}
\includegraphics[width=3.5in]{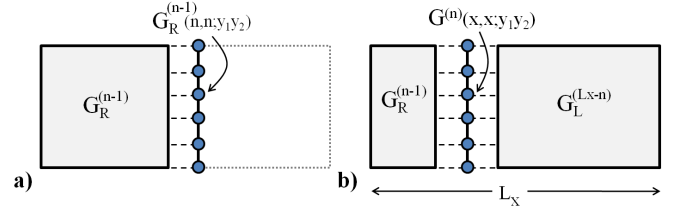}
\end{center}
\vspace{-.2in}
\caption{(Color online) Schematic depiction of recursive Green's function method (eq. \ref{eq:GrnFnMethod1},\ref{eq:GrnFnMethod2}).  (a) Green's function for the end of an $L_x$-layer sample $G^{(L_x)}$ is calculated recursively from $G^{(n-1)}\rightarrow G^{(n)}$ by adding one transverse layer at a time.  (b) Green's function for any layer can then be recovered from solving (\ref{eq:GrnFnMethod2}) for the desired location, using the intermediate results for $G^{(L_x-x)}$ and $G^{(x-1)}$ from the top procedure.}
\label{fig:GrnFnMethod}
\end{figure}

For sufficiently small system size, the full 2D tight-bonding model (\ref{eq:H_TB}) can be analyzed directly by exact diagonalization.  For larger systems, where exact diagonalization becomes computationally intractable, we employ a recursive Green's function method\cite{LeeFisher} depicted schematically in Fig. \ref{fig:GrnFnMethod}.  Starting with the Green's function for the right (left) end of a strip of length $n$, $G_{R(L)}^{(n)}(y,y')$ one can construct the Green's function for the right (left) end of a strip of length $n+1$ by:
\begin{eqnarray} G_R^{(n+1)}(\omega) &=& \frac{1}{\omega+i\eta-H_{n}-V^\dagger G_R^{(n)} V} \nonumber \\
G_L^{(n+1)}(\omega) &=& \frac{1}{\omega+i\eta-H_{n}-V G_L^{(n)} V^\dagger} \label{eq:GrnFnMethod1}\end{eqnarray}
where $H_{n}$ is the Hamiltonian for the strip at location $x=n$, $V$ is the matrix containing all elements of (\ref{eq:H_TB}) that connect $x=n$ to $x=n+1$, and $\eta>0$ gives the poles of $G$ a small imaginary component.  By repeated recursion of (\ref{eq:GrnFnMethod1}), one can find the Green's function for the end of an arbitrarily long strip.  The Green's function for any fixed value of $x$ can then be calculated from:
\begin{eqnarray} && G^{(L_x)}(x,x;y_1,y_2;\omega) = \nonumber \\
 && \hspace{.4in}\frac{1}{\omega+i\eta-H_{x}-V^\dagger G_R^{(x-1)} V-V G_L^{(L_x-x)} V^\dagger} \label{eq:GrnFnMethod2} \end{eqnarray}
Once calculated, the Green's function can yield the density of states: $\rho(\omega) = -\frac{1}{\pi}\text{Tr}\Im m G(\omega)$, which has a $\delta$-function peak at each energy level.  Furthermore, for $\omega\rightarrow\e_n$, where $\e_n$ is an eigen-energy of (\ref{eq:H_TB}), the Green's function becomes a projector onto the corresponding eigenstate $\Psi_n(\mathbf{R})$: $\lim_{\omega\rightarrow \e_n} (i\eta)G(\mathbf{R},\mathbf{R};\omega) \rightarrow |\Psi_n(\mathbf{R})|^2$.

\section{Hard-Wall Confinement}
We first consider the the tight-binding Hamiltonian (\ref{eq:H_TB}) on a long strip of length $L_x$ and width $L_y$ with hard-wall boundaries.  These boundary conditions correspond to strong confinement and are relevant, for example, to self-assembled semiconductor nano-wires or etched semiconductor or metallic microstrips.  

\begin{figure}[ttt]
\begin{center}
\hspace{-.2in}
\includegraphics[width=3.5in]{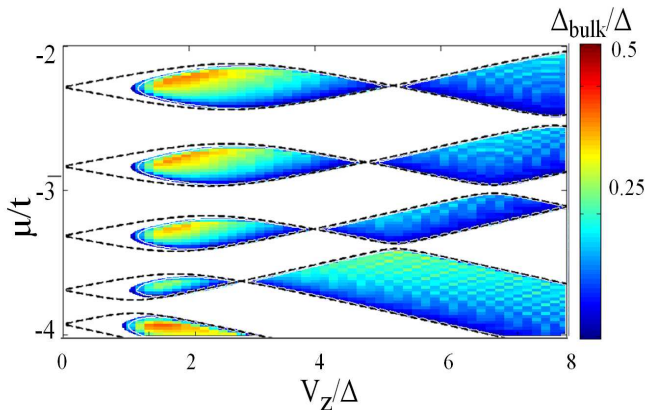}
\end{center}
\vspace{-.2in}
\caption{(Color online) Phase diagram as a function of chemical potential $\mu$ and Zeeman splitting $V_z$ for strip of width $L_y=10$, with hard-wall boundaries and $t=10$, $\Delta=1$, and $\alpha_R=2$.  White regions correspond to the topologically trivial phase with no Majorana end-states.  Colored regions indicate topologically non-trivial phases with Majorana end-states, color indicates size of bulk excitation gap $\Delta_{\text{bulk}}$ which sets the length scale $\ell_M \sim v_F/\Delta_{\text{bulk}}$ over which the end-states decay along the length of the wire.  Note the reduction of $\Delta_{\text{bulk}}$ as $V_z$ becomes large.  Dashed lines show the location of the non-superconducting ($\Delta=0$) transverse sub-bands, which are closely related to the location of topological phase-transitions.}
\label{fig:HardWallPD1}
\end{figure}

The bulk excitation gap, $\Delta_{\text{bulk}}$, for an infinitely long strip of width $L_y$ provides a convenient way of characterizing topological phase transitions. The locations of topological phase transitions, corresponding to closings of the bulk-gap, can be simply obtained by solving the effectively 1D problem of diagonalizing (\ref{eq:H_TB}) at zero momentum along x ($k_x=0$).  Using the Green's function method outlined above, we verify that closings of the bulk gap ($\Delta_{\text{bulk}}=0$) in an infinitely long strip signify topological phase transitions in a finite length strip.  These transitions are between states with zero-energy Majorana end-modes, and topologically trivial states without zero-energy end-modes.  Fig. \ref{fig:HardWallPD1} shows the topological phase-diagram as a function of chemical potential $\mu$ and Zeeman splitting $V_z$ for a strip of width $L_y=10$ with parameters $t=10$, $\Delta=1$, and $\alpha_R=2$.   Fig. \ref{fig:HardWallPD1} is compatible with the results of Ref.  \onlinecite{DasSarmaMultichannel}, which computes the phase diagram restricted to the lowest two sub-bands.

 The effect of transverse confinement in the y-direction is to quantize y-momenta $k_y$ to discrete values: $k_n$ with $n\in\{1,2,\dots\}$, each with a corresponding a 1D transverse sub-band.  In Fig. \ref{fig:HardWallPD1}, the transverse sub-band bottoms for the infinite strip with $\Delta=0$ are shown as dashed lines.  These transverse sub-bands are doubly spin-degenerate for $V_z=0$, and are split linearly as $V_z$ is increased.  Without Rashba splitting, this would result in a fan of crossing levels as $V_z$ is increased, however the presence of Rashba splitting leads to avoided crossings between the different sub-bands.

Similarly to the case of multi-band of spinless $p+ip$ superconducting quantum wires\cite{Potter}, we find that zero-energy Majorana end-states exist only when an odd-number of these transverse sub-bands are occupied.  However, the picture here is complicated by the presence of additional energy scales $\alpha_Rk_F$ and $V_z$ which give rise to two spin-split surface bands with energies $\e_{\pm} = \xi_k\pm\sqrt{V_z^2+\alpha_R^2k^2}$.  As described above, pairing between two electrons of the same helicity has $p_x\mp ip_y$ symmetry and amplitude $\Delta_{\text{p}}$, whereas pairing between electrons of different helicity has s-wave symmetry with amplitude $\Delta_s$ (see eq. \ref{eq:DeltaSP}).  When $V_z$ is small enough that $\e_{+}(k_n)-\e_{-}(k_n)<\Delta$, both $p+ ip$ and $p-ip$ pairing occurs and the system is topologically trivial for all $\mu$.  In contrast, for sufficiently large $V_z$ such that $\e_{+}(k_n)-\e_{-}(k_n)>\Delta$ only one of the two $p\pm ip$ components remains. In this regime, so long as the quantized levels of $\e_{\pm}(k_y=k_n)$ are offset, there are intervals of $\mu$ for which an odd number, $\mathcal{N}$ of occupied sub-bands. In such intervals there are pairs of $p+ip$ and $p-ip$ bands, and one unpaired $p+ip$ band.  One can picture each sub-band as contributing a single Majorana end-state; these are intercoupled and mix, to form $\lfloor\(\mathcal{N}/2\)\rfloor$ full fermions at non-zero energy and $\mathcal{N}\text{mod}2$ Majorana zero-modes.

By these considerations alone, it would seem that larger $V_z$ is always favorable for creating Majorana end-states.  However, the gap protecting the Majorana end-states from bulk excitations depends on the p-wave component $\Delta_p$ of the induced pairing gap, which for $V_z\gg m\alpha_R^2$ scales as: $\Delta_{\text{bulk}}\simeq \Delta_p \rightarrow \frac{m\alpha_R^2}{V_z}\Delta \ll \Delta$ (see Eq. \ref{eq:DeltaSP}).  Consequently, there is a trade off between increasing $V_z$ to stabilize Majorana end-states, and in avoiding $V_z\gg \alpha_R$ to protect the bulk pairing gap.

\begin{figure}[ttt]
\begin{center}
\hspace{-.2in}
\includegraphics[width=3.5in]{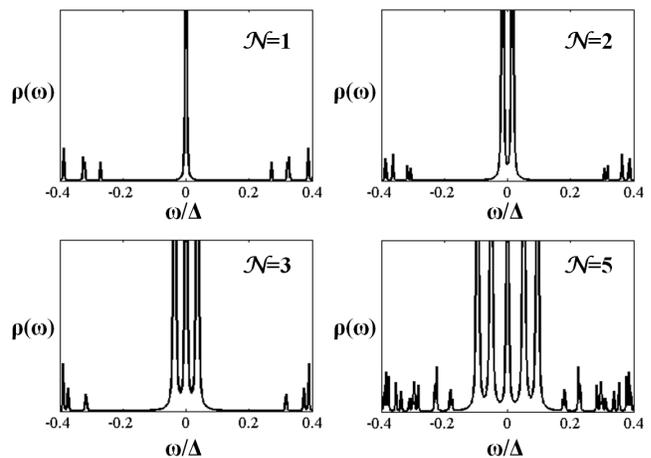}
\end{center}
\vspace{-.2in}
\caption{Spectral function for the end layer of a $200\times 10$ strip for the tight-binding model (\ref{eq:H_TB}) with hard-wall boundaries as determined by Green's function method (\ref{eq:GrnFnMethod1}) with $\eta/\Delta=0.001$.  Simulation parameters used were: $t=10$, $\Delta=1$, $\alpha_R=2$, and $V_z = 6$.  From left to right and top to bottom, the chemical potentials are $\mu/t = $ -4.37, -4.09, -3.72, and -3.265 corresponding to $\mathcal{N}=$ 1,2, 3 and 5 transverse sub-bands occupied respectively.  There is a zero-energy pole only for $\mathcal{N}$ odd.  For general $\mathcal{N}$ there are $\mathcal{N}$ end-states split by the effective p-wave gap $\sim \Delta_{\text{p}}$.}
\label{fig:DOS}
\end{figure}

This picture of each sub-band contributing a Majorana end-state is supported by Fig. \ref{fig:DOS}a.--d. which show the spectral function for a $L_x=200$, $L_y=10$ strip, projected onto the last layer.  Fig. \ref{fig:DOS}a.--d. were simulated using the same parameters as for Fig. \ref{fig:HardWallPD1} with fixed $V_z=6$, and $\mu$ adjusted such that $\mathcal{N}=$ 1, 2, 3, and 5 transverse sub-bands are occupied respectively.  In each subfigure, one finds two types of excitations.  The first are bulk excitations above $\Delta_{\text{bulk}}\sim 0.25\Delta$.  These reside along the length of the strip and have small projection  $\sim 1/L_x$ onto the end-layer.  The second are end-states.  The end state wave-functions are exponentially localized to the sample ends, with localization length $\xi_M\sim v_F/\Delta_{\text{bulk}}$ set by the bulk excitation gap.  Consequently, these states have large projection onto the end-layer that is independent of $L_x$ for $L_x\gg \xi_M$.  As shown in Fig. \ref{fig:DOS} for $\mathcal{N}$ occupied sub-bands, there are $\mathcal{N}$ end-states, and we have verified that they are roughly equally spaced in energy $\sim\Delta_p/\sqrt{\mathcal{N}}$.  For $\mathcal{N}$ odd, one of these end states sits at exactly zero energy, and is a Majorana state protected from excitations by energy gap: $\sim\Delta_{\text{p}}$.  For $\mathcal{N}$ even the end-states sit at non-zero energy and are not Majorana fermions.

To summarize, we find that, in hard-wall confined strips, Majorana end states exist as long as the $\e_+-\e_->\Delta$ and so long as there are an odd number of occupied transverse sub-bands.  Surprisingly, this alternating even-odd behavior persists in confined systems even for $|\mu|>V_z$, where the 2D bulk material would be \emph{non-topological}.  This resilience is due to the fact that the bulk $p+ip$ and $p-ip$ subbands are offset by different quantization energies.  

These results offer two principal advantages for realizing Majorana fermions as end states in multi-band confined structures.  First, the chemical potential in candidate materials may naturally lie far away from $\mu=0$.  For non-confined schemes, one would need to shift $\mu$ to lie in the Zeeman gap, possibly requiring chemical doping or large electrostatic gate voltages.  Here, one only needs to fine-tune the chemical potential on the order of $E_{\text{sb}}\sim \frac{\pi^2}{2mL_y^2}$. Secondly, as discussed in Ref. \onlinecite{AliceaSingle}, because of the extremely small Rashba splittings available in semiconductor nanowires, restricting $|\mu|<V_z$ in these materials would require operating at exceedingly small carrier density.  However, the above results demonstrate that one may work far outside this $|\mu|<V_z$ regime, opening the door to substantially higher densities.  

Despite this, small Rashba couplings still pose a serious problem for semiconductor materials.  In order to remain in the topological regime, one needs $V_z>\Delta$.  Given that typical values of $m\alpha_R^2$ in semiconductor materials are of the order of $0.2$K\cite{AliceaSingle}, and assuming that $V_z\sim\Delta$ is on the order of a few Kelvin, one has $V_z/m\alpha_R^2\lesssim 1/10$.  This means that only a small fraction of the superconducting gap would be converted into p-wave pairing, requiring one to operate at very low temperatures in order to avoid thermal excitations.

\section{Ferromagnetic Domains}
We next consider a fully two-dimensional version of (\ref{eq:H_TB}) in which $V_z$ is non-zero only inside a long finite strip of length $L_x$ and width $L_y$.  This could be accomplished by depositing a narrow  strip of FM insulator on a fully 2D Rashba coupled surface.  The idea here is that, under the right conditions, the ferromagnetic strip creates a strip of topological $p+ip$ superconductor embedded in a topologically trivially $p\pm ip$ superconducting background.  Such spatial interfaces between regions with different topological ordering generally give rise to localized zero-modes.

\begin{figure}[bbb]
\begin{center}
\hspace{-.2in}
\includegraphics[width=3in]{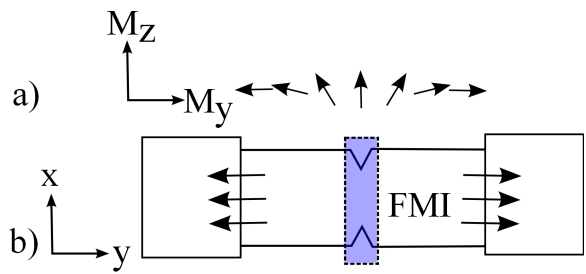}
\end{center}
\vspace{-.2in}
\caption{(Color online) A smoothly varying ferromagnetic domain wall can be created by creating a notch in a strip in an easy-plane ferromagnetic insulator (FMI), and seeding the opposite sides of the notch with opposite (in-plane) magnetization. Panel b shows a top view of the proposed notch geometry, with arrows indicating the magnetization seed directions.  Panel A shows a side view of the resulting magnetization profile (y- and z- components of the magnetization $M$), which tips preferentially into the z-direction at the location of the notch, creating a thin region of non-vanishing z-axis polarization corresponding to the shaded blue region in b.}
\label{fig:FMDomainNotch}
\end{figure}

Figure \ref{fig:FMDomainNotch} shows a convenient geometry for realizing such a narrow domain wall.  In this setup, the ferromagnetic insulator (FMI) layer has boundary conditions such that the magnetization points in the $-y$ direction on the leftmost side and in the $+y$ direction on the rightmost side of the film.  A narrow notch in the center of the film seeds the domain wall in the center of the sample.  Since thin films of FMI are typically easy-plane materials (i.e. the magnetization prefers to lie in the x-y direction), it is energetically favorable for the domain to form by rotating the magnetization out of the plane along the z-axis.  Since only the $z$-direction magnetization opens up a Zeeman gap in the underlying Rashba coupled material, this domain wall creates a narrow topologically non-trivial region (shown as a shaded blue rectangle in panel b. of Figure \ref{fig:FMDomainNotch}).  The resulting magnetization profile automatically varies smoothly with position, avoiding rough edges which would create pair breaking scattering.

\begin{figure}[bbb]
\begin{center}
\hspace{-.2in}
\includegraphics[width=3.5in,height=3in]{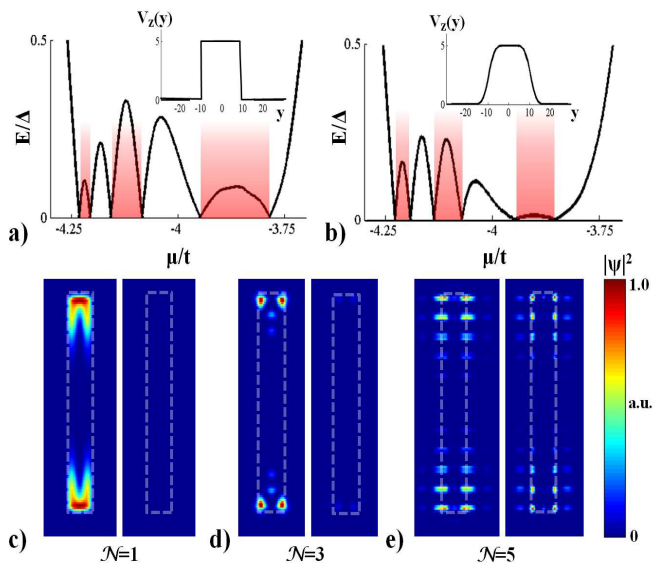}
\end{center}
\vspace{-.2in}
\caption{(Color online) (Top) Bulk excitation spectrum for infinitely long ferromagnetic (FM) strip-domain of width $L_y$ with sharp edges (a) and Gaussian-smoothed edges (b). Highlighted red regions denote the presence of zero-energy end states as determined by the Green's function method (\ref{eq:GrnFnMethod1}) for strips of length $L_x=100$.  Insets show $V_z(y)$ profiles.  Tight-binding parameters used were $t=20$, $\alpha_R=10$, $V_z=5$, and $L_y=20$.  (Bottom) Corresponding wave-function $|\Psi|^2$ profiles for finite length ($L_x=100$) strips with sharp $V_z$ profile as in (a).  Subfigures (c),(d), and (e) show $|\Psi|^2$ for selected chemical potentials $\mu/t =$ $-4.22$, $-4.12$, and $-3.87$, which correspond to $\mathcal{N}=1,3$ and $5$ occupied sub-bands respectively.  Spin-down (left) and spin-up (right) components shown separately.  Faint dashed-lines mark the boundary of the FM strip.}
\label{fig:VzStrip}
\end{figure}

Fig. \ref{fig:VzStrip}a. shows the bulk excitation gap as a function of chemical potential $\mu$ for an infinitely long FM strip of width $L_y=20$ embedded in a large non-FM coupled background with parameters $t=20$, $\Delta=10$, $\alpha_R=10$, and $V_z=5$.  In this simulation $V_z$ has a sharp step profile, and drops abruptly from $V_z=5$ inside the strip to $V_z=0$ outside.  Red shaded regions indicate the presence of Majorana end-states as determined using the Green's function method (\ref{eq:GrnFnMethod1}).  As with the case of hard-wall boundary conditions, we find an alternating sequence of topological phase transitions as $\mu$ is varied.  These transitions again occur where the bulk gap for an infinitely long strip closes. However, unlike the hard-wall boundary case considered above, $k_y$ is no longer quantized by confinement and the bulk gap closings are not determined by transverse momenta quantization.  Rather, the spatially inhomogenous $V_z$ profile gives rise to an effective step potential which binds discretely spaced 1D energy bands, whose bottoms lie within the bulk superconducting gap.  Once again, Majorana end-states appear when an odd number of these discrete energy bands are occupied. 

Fig. \ref{fig:VzStrip}c.--e. show wavefunction profiles of the spin-up and down components of the Majorana end-states for $\mathcal{N}$ = 1, 3, and 5 occupied subbands. The wave-functions are exponentially localized to the ends of the FM strip, with characteristic localization length set by the bulk excitation gap.  The number of nodes in the end-state wave-function increases with $\mathcal{N}$.  Furthermore, since the $V_z$ potential step profile favors spin-down, the more deeply confined $\mathcal{N} = $ 1 and 3 states have spin-down, the shallowly confined $\mathcal{N}=5$ band has a more equal mixture of spin-up and down.

In realistically fabricated structures where $V_z$ is induced by proximity to a FM insulator, the $V_z$ profile will likely not be a sharp step function.  Fortunately, we find that the above picture is largely insensitive to the details Zeeman splitting spatial profile $V_z(y)$. For example, Fig. \ref{fig:VzStrip}b. repeats Fig. \ref{fig:VzStrip}a. with smoothly varying $V_z(y)$, obtained by applying a Gaussian filter with width of 3 lattice spacings to a sharp step-profile.  The alternating phase transition structure and presence of Majoranas is very similar to the sharp step-profile case.  The main difference being that the larger $\mathcal{N}$ sub-bands are more shallowly confined by the smoother $V_z$ potential.  

We have shown that the existence of Majorana end states in this setup is robust, persisting so long as the FM domains are narrow, elongated structures with gapped mid-sections. These results make it feasible to produce Majorana states in FM domains patterned on top of bulk 2D Rashba split surfaces with induced superconductivity.   

\section{Majorana End-Modes Without Transverse Sub-bands}
\subsection{Random Edge Geometries}
In the long rectangular strip geometries considered above, the system is neatly separable in the x and y directions.  So far, the existence of discrete transverse sub-bands (in the y-direction) has played a central role in understanding the topological phase transitions in these structures.  Naturally, one might therefore wonder whether the existence of transverse sub-bands is essential to the formation of Majorana end-states.  Specifically, the presence of spatially varying and non-parallel edges mixes different transverse sub-bands, destroying the notion of the ``number of occupied channels".  Since, for rectangular samples, Majorana end-states exist only for an odd-number of transverse channels, it is possible that the mixing of even and odd number of channels may destroy the Majorana end-states in non-rectangular samples.

To address this question, we consider samples confined to a narrow region by electrostatic confinement potential $V_{\text{conf}}(x,y)$ with smooth, random meandering boundaries. 
Here it is important that the edge variation is relatively smooth, as jagged edge variations produce a scattering mean-free path $\ell \simeq W$. Due to the condition $W\lesssim \xi_0$, scattering from sharp edge variations tends to destroy the p-wave pairing gap\cite{Potter}.  

To produce random edges with width variance $\sigma_W$ and correlation length $\ell$, we start by choosing the y-location of the top and bottom edges $y_{t,b}(x)$ independently for each $x$, identically distributed normally with variance $\sqrt{2\sigma_W^2\ell}$ and mean $\overline{L}_y/2$ (where the over-bar indicates averaging with respect to edge configuration).  We then apply an exponential smoothing filter $y_{t,b}(x)\rightarrow \sum_{x'} \frac{1}{\ell} e^{-|x-x'|/\ell}y_{t,b}(x')$, which correlates $y_{t,b}(x)$ and $y_{t,b}(x')$ on lengthscales, $|x-x'|\lesssim \ell$, on the order of the edge-correlation length $\ell$.  

\begin{figure}[ttt]
\begin{center}
\hspace{-.2in}
\includegraphics[width=3.2in]{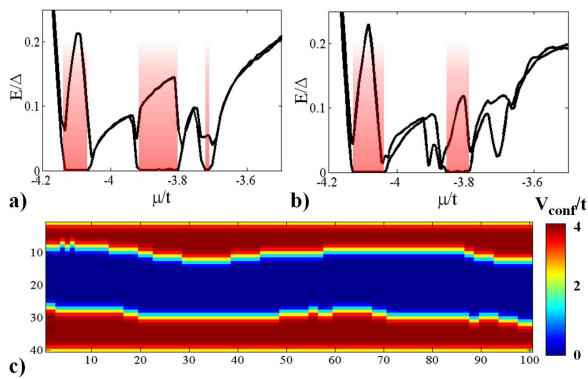}
\end{center}
\vspace{-.2in}
\caption{(Color online) (a) and (b) show the two lowest energy in-gap excitations for an electrostatically confined strip with $L_x=100$, $\overline{L}_y=20$, $t=10$, $\Delta=1$, $V_z=2$, $\alpha_R=2$.  Red shading indicates the presence of isolated Majorana end-modes at zero-energy.  The results in (a) are for straight edges ($\sigma_W=0$), and those  in (b) are for a random sample with $\sigma_W=4$ and $\ell=15$; (c) shows a colormap of the random edge geometry used to generate (b).  Importantly the Majorana edge states survive, retaining a substantial excitation gap even for large edge variation (in this case $\sim 40\%$ of the average width $\overline{L}_y$) and demonstrating that these states do not rely on the existence of transverse sub-bands.}
\label{fig:RandomEdge}
\end{figure}

\begin{figure}[ttt]
\begin{center}
\hspace{-.2in}
\includegraphics[width=3in]{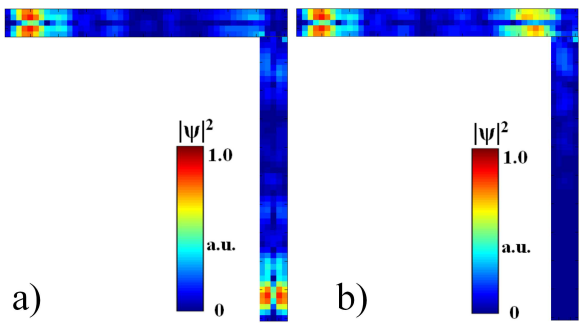}
\end{center}
\vspace{-.2in}
\caption{(Color online) Spatial profile of the Majorana wave-function intensity, $|\Psi(x,y)|^2$, for L-shaped junction, for $t=10$, $\Delta=1$, $\alpha_R=5$, and $V_z=2$.  The dimensions of each leg of the junction are $5\times 50$ lattice sites.  The left (a) shows the wave-function with three occupied sub-bands in each leg ($\mu=-3t$).  In this case the Majorana states exist at the extreme ends of the the L. The right (b) shows the wave-function with three sub-bands  occupied in the horizontal leg ($\mu=-3t$), and two sub-bands occupied in the vertical leg ($\mu=-3.4t$).  In this case the second Majorana mode appears at the junction between topological and non-topological regions at the elbow of the L-junction.}
\label{fig:LJunction}
\end{figure}

Since the sample width at any $x$ must be an integer number of lattice spacings, $\{y_{t,b}(x)\}$ are rounded to the nearest integer, resulting in discrete steps rather than smooth edges.  These steps introduce sharp, short range scattering potentials, and in order to separate out the effects of this discretization, from those of the smoothly wandering edges, we smooth the $V_{\text{conf}}$ along the lateral (y--) direction with a Gaussian filter of width 2 lattice spacings.    

Fig. \ref{fig:RandomEdge}a. and b. show the results of simulations with a smoothly random electrostatic confinement potential for samples with average width $\overline{L}_y=20$, length $L_x = 100$, edge correlation length $\ell = 15$, and with edge variance $\sigma_W = 0$ and $4$ respectively.  We find that the Majorana end modes, and corresponding sequence of alternating phase transitions survives even for substantial edge variations, that is, even when there are no well-defined transverse sub-bands.  Despite the lack of transverse sub-bands, as one sweeps $\mu$, discrete bulk levels inside the superconducting gap are still pulled down one-by-one across zero-energy, resulting again in an alternating sequence of topological phase transitions.  However, the locations of these transitions occur at different values of $\mu$ compared to the rectangular case.  In contrast to the rectangular sample case, these discrete levels cannot be simply identified with transverse band-bottoms, but rather are bulk states with some more complicated structure.  The excitation gap protecting Majorana end-states in these random edge geometries is reduced from the rectangular case.  However, as seen by comparing Fig. \ref{fig:RandomEdge}a. and b., this excitation gap remains a substantial fraction of the straight edge rectangular case even for large variations in the edge geometry (in the case of Fig. \ref{fig:RandomEdge} b. the fractional variation in width, $(\overline{\delta y_t^2}+\overline{\delta y_b^2})/\overline{L}_y = 2\sigma_W/\overline{L}_y$, is $40\%$).

These simulations demonstrate that the existence Majorana end-states is highly insensitive to the details of sample geometry, and in particular \emph{does not require the existence of transverse sub-bands}.  This robustness to edge-variations highlights the truly topological nature of these states.  Also from a practical perspective, the ability to tolerate substantial (smooth) edge variance eases the requirements for sample fabrication, making an experimental realization more feasible.

\subsection{L-shaped Junction}
Majorana end-states can be moved, braided, and fused using appropriate networks of quantum wires\cite{Alicea1DWires}.  A common feature of such networks are right-angle junctions (either L- or T- shaped).  Fig. \ref{fig:LJunction} provides an explicit demonstration that the Majorana end-states with mulitply occupied sub-bands, even in geometries with sharp corners.  Fig. \ref{fig:LJunction} a) shows the wave-function for an odd-number of occupied sub-bands  in both the horizontal and vertical legs of an L-junction, whereas Fig. \ref{fig:LJunction} b) shows the wave-function for an odd-number of occupied sub-bands in the horizontal and leg and an even number in the vertical leg.  In the former case, the entire L-junction is topological, and Majorana end-states reside at the extremal ends, whereas in the latter case, the Majorans lie entirely within the horizontal leg, with the second Majorana occuring at the junction between topological and non-topological regions. 

\section{Conclusion and Discussion}
In conclusion, by numerically diagonalizing the tight-binding model \ref{eq:H_TB}, we have shown that quasi-one-dimensional microstructures with Rashba spin-orbit coupling are are a robust medium in which to realize Majorana fermions.  In spatially confined structures, such as semiconductor nanowires, confinement quantization enables one to operate at substantially higher carrier densities where a bulk 2D material would be non-topological.  Additionally, the presence of multiple energy scales allows one to trap Majorana fermions at the ends of quasi-one-dimensional ferromagnetic domains, or in electrostatically confined strips.  

Importantly, Majorana end-states realized in this way are largely immune to both bulk disorder\cite{Potter} and random sample geometry (so long as the edges are relatively smooth).  Furthermore, for end-states, the mini-gap to localized excitation scales as $\Delta$, rather than $\Delta^2/\e_F$ as for vortex core states.  While it has been argued that minigaps are irrelevant for detection schemes based solely on electron number parity\cite{VortexInterferometry} (in which Majoranas are kept far apart from each other), our results show that Majorana end-states allow for simpler fusion based measurement schemes.  Finally, the ability to selectively tune segments of these systems through a topological phase transition, simply by electrostatic gating, provides a convenient, scalable avenue towards manipulating and braiding many Majorana fermions.  

\textit{Acknowledgements - } We thank A.R. Akhmerov, A. Kitaev, and J.S. Moodera for helpful discussion. This work was supported by DOE Grant No. DE--FG02--03ER46076 (PAL) and NSF IGERT Grant No. DGE-0801525 (ACP).


\begin{thebibliography}{}

\bibitem{Potter} 
A.C. Potter and P.A. Lee, Phys. Rev. Lett. {\bf 105}, 227003 (2010).
\bibitem{Read/Green}
N. Read and D. Green, Phys. Rev. B {\bf 61}, 10267 (2000).
\bibitem{Ivanov}
D.A. Ivanov, Phys. Rev. Lett. {\bf 86}, 268 (2001).
\bibitem{Nayak}
C. Nayak, S.H. Simon, A. Stern, M. Freedman, and S. Das Sarma, Rev. Mod. Phys. {\bf 80}, 1083 (2008).
\bibitem{Kitaev}
A. Kitaev, arXiv:cond-mat/0010440 (2000).
\bibitem{He3Ref}
Lee, D. M. Rev. Mod. Phys. 69, 645–665 (1997). 
\bibitem{Sr2RuO4Ref}
G. M. Luke et al. Nature {\bf 394}, 558-561 (1998); K. Ishida et al. Nature {\bf 396}, 658-660 (1998) 
\bibitem{Moore/Read}
G. Moore and N. Read, Nuc. Phys. B {\bf 360}, 362 (1991).
\bibitem{Fujimoto}
S. Fujimoto, Phys. Rev. B {\bf 77}, 220501 (2008).
\bibitem{Sau}
J.D. Sau, R.M. Lutchyn, S. Tewari and S. Das Sarma, Phys. Rev. Lett. {\bf 104}, 040502 (2010).
\bibitem{AliceaSingle}
J. Alicea, Phys. Rev. B {\bf 81}, 125318 (2010).
\bibitem{PALee}
P.A. Lee, arXiv: 0907.2681.
\bibitem{DasSarmaMultichannel}
R.M. Lutchyn, T. Stanescu, S. Das Sarma, arXiv:1008.0629 (2010).
\bibitem{VortexInterferometry}
F. Hassler, A. R. Akhmerov, C.-Y. Hou, C. W. J. Beenakker, arXiv:1005.3423 (2010).
\bibitem{Fu/Kane}
L. Fu and C.L. Kane, Phys. Rev. Lett. {\bf 100}, 096407 (2008).
\bibitem{Sau2}
J.D. Sau, R.M. Lutchyn, S. Tevari, and S. Das Sarma, arXiv: 0912.4508

\bibitem{LeeFisher}
P.A. Lee and D.S. Fisher, Phys. Rev. Lett. {\bf 47}, 882 (1981).
\bibitem{Alicea1DWires}
J. Alicea, Y. Oreg, G.l Refael, F. von Oppen, M.P.A. Fisher,	arXiv:1006.4395 (2010).
\end{thebibliography}
\end{document}